\documentclass[aps,prb,twocolumn]{revtex4}
\usepackage{epsfig}
\begin{document} 
\title{Carrier-induced ferromagnetism in a diluted Hubbard model}
\author{Sudhakar Pandey and Avinash Singh}
\email{avinas@iitk.ac.in} 
\affiliation{Department of Physics, Indian Institute of Technology Kanpur - 208016}
\begin{abstract}
Carrier-induced ferromagnetism is investigated in a diluted Hubbard model
for ordered impurity arrangements.
The delicate competition between particle-hole processes contributing
to the spin couplings results in a rich variety of behaviour.
The ferromagnetic transition temperature obtained within the
spin-fluctuation theory is in good agreement with
reported values for $\rm Ga_{1-x} Mn_x N$.
\end{abstract}
\maketitle
\section{Introduction}
The recent discovery of ferromagnetism in diluted magnetic semiconductors (DMS) such as
$\rm Ga_{1-x} Mn_x As$\cite{GaMnAs1,GaMnAs2} and $\rm Ga_{1-x} Mn_x N$,\cite{GaMnN3}
and the intensive efforts to increase the ferromagnetic 
transition temperature in view of potential technological applications, 
has generated tremendous interest in the novel ferromagnetism exhibited by these systems
in which magnetic interaction between localized spins is mediated by doped 
carriers.\cite{mf1,mf2,akai,mf3,mf4,mf6,mf7,dmft1,dmft2,non,mc1,mc2,mc3,prelim,squid,timm}
The nature of the ferromagnetic state has also attracted much attention,
particularly in the context of competing antiferromagnetic (AF) interaction
which result in noncollinear ordering, spin-glass behaviour,
and significant sensitivity of spin stiffness and transition temperature $T_c$
on carrier concentration, spin clustering etc.

DMS are mixed spin-fermion systems, involving randomly distributed localized magnetic impurities
(e.g., Mn$^{++}$, having $S=5/2$) and mobile carriers (e.g., holes) in the semiconductor band. 
With carrier concentration $p$ much smaller than the magnetic impurity concentration $x$, 
the DMS systems provide a complimentary limit to Kondo systems.
Conventionally, the coupling between localized impurity spin ${\bf S}$ 
and mobile valence band holes is represented by the exchange interaction 
$-J {\bf S}. \mbox{\boldmath $\sigma$} $,
where $\mbox{\boldmath $\sigma$} = \Psi^\dagger
[\mbox{\boldmath $\sigma$}] \Psi$ is the fermion spin operator.  

Cuprates form another class of strongly correlated systems
in which the concentration of doped carriers is small in comparison
to that of (Cu) spins.
Extensively studied within the Hubbard model,
the dominant interaction in cuprates is the AF exchange interaction
between neighbouring Cu spins,
and AF spin correlations persist even for small hole doping.
Against this strong tendency for AF ordering,
the delocalization energy gain of doped carriers,
which favours parallel spin alignment,
results in only a marginal twisting of spins,
as in the proposed spiral phases.\cite{spiral1,spiral2}
It is therefore interesting that elimination of
the strong AF spin interaction in a {\em diluted} Hubbard model,\cite{prelim}
with no direct hopping between relatively distant impurity spins,
does indeed lead to a ferromagnetic state stabilized by
carrier-induced spin couplings.

The spin stiffness in the ferromagnetic state of the diluted Hubbard model
goes through a maximum with respect to both doping concentration $p$
and the interaction strength $U$.
This optimization behaviour can be qualitatively understood in terms
of a competition between the increasing magnitude of
carrier spin polarization $\chi(U)$ and the increasing rapidity of its
oscillation, which limits the growth of the spin coupling $U^2 \chi_{ij}(U)$
between two magnetic impurities at a fixed separation.
Similar behaviour was observed in the spin-fermion model
for the effective ferromagnetic coupling $J_{ij}=J^2 \chi_{ij}(J)$
in terms of the generalized spin response $\chi(J)$ for finite $J$.\cite{prelim}
As $\chi(J)$ involves particle-hole processes, 
the spin couplings identically vanish in the absence of doping.
However, in the diluted Hubbard model,  
even with no carrier doping ($p=0$) in the majority-spin band,
particle-hole processes involving the
empty minority-spin impurity band result in 
antiferromagnetic spin couplings,
which destabilize the ferromagnetic state below a critical
doping concentration.

In this paper, we study the novel ferromagnetic state of the diluted
Hubbard model for {\em ordered} impurity arrangements.
We focus on magnon excitations and the spin stiffness, 
which provide quantitative measures of
the stability and spin couplings in the ferromagnetic state,
as well as the transition temperature in three dimensions.
While preliminary real-space studies for both ordered and disordered impurity
arrangements were carried out earlier on three-dimensional finite-size
lattices,\cite{prelim}
use of ${\bf k}$-space representation in this paper allows for 
much larger lattices, thus permitting a more refined study of the
competing spin interactions with respect to
carrier concentration, impurity separation, interaction strength, and wave vector.
Indeed, the antiferromagnetic - ferromagnetic quantum phase transition
stands out as a significantly prominent feature in the low doping regime.
AF-F and spin-glass transitions at low doping have actually been observed
in Mn-doped II-VI semiconductors.\cite{af-f1,af-f2,sg1}
With increasing doping concentration,
an anomalous increase in the spin stiffness is observed
which involves, as explained in section III B, a subtle interplay of
impurity moment reduction, impurity character of doped states, and
competing particle-hole processes,
all characteristics of the itinerant ferromagnetic state. 

Within the relatively simpler ferromagnetic Kondo lattice model (FKLM),
in which a localized spin is present at every lattice site,
magnon excitations have been studied recently, as a function of electron
density $n$ in the conduction band and the spin-fermion coupling $J$,
in the context of heavy fermion materials,\cite{sigrist+ueda}
ferromagnetic metals Gd, Tb, Dy, doped EuX,\cite{donath}
and manganites.\cite{furukawa,wang,vogt}
Magnon dispersion has also been obtained
in the context of DMS.\cite{koenig,pareek,dis4,dms_nolting} 
However, a uniform impurity-induced Zeeman splitting of the carrier
spin bands is assumed.\cite{koenig,pareek,dms_nolting} 

\begin{figure}[hbt]
\includegraphics[angle=0,width=.4\columnwidth]{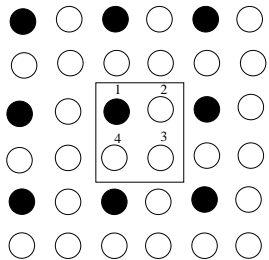}
\caption{An ordered arrangement of impurity atoms ($\bullet$)
on a square host lattice ($\circ$). 
Also shown are the sublattice labels corresponding to the four distinct
sites within the unit cell.}
\end{figure}

\section{Diluted Hubbard model}
We consider a diluted Hubbard model 
\begin{eqnarray}
H &=& t \sum_{i,\delta,\sigma} a_{i,\sigma}^{\dagger}a_{i+\delta,\sigma}
+ t' \sum_{I,\delta,\sigma} a_{I,\sigma}^{\dagger}a_{I+\delta,\sigma}
\nonumber \\
&+& \epsilon_d \sum_{I,\sigma} a_{I\sigma}^\dagger a_{I\sigma}
+ U \sum_{I} 
\left (n_{I\uparrow}- \langle n_I \rangle \right )
\left (n_{I\downarrow}- \langle n_I \rangle \right )
\nonumber \\
\end{eqnarray}
on square and cubic lattices with nearest-neighbour (NN) hopping
between sites $i$ and $i+\delta$.
Here $I$ refers to the impurity sites,
$\epsilon_d$ is the impurity on-site energy,
and $\langle n_I \rangle= \langle n_{I\uparrow} + n_{I\downarrow} \rangle /2$
is the spin-averaged impurity density. 
The energy-scale origin is set so that the host on-site energy is zero,
and we take the impurity level to lie at the top of the host band
($\epsilon_d = 2dt =W/2$ in $d$ dimensions) to optimize local-moment formation.
For a positive sign of the hopping term,
the top of the host band lies at ${\bf k}=0$, as in the valence band of DMS.
For simplicity, we take the same hopping ($t'=t$) 
between the host-host and host-impurity pairs of sites.
Higher spin magnetic impurities, such as the $S=5/2$ Mn impurities in 
$\rm Ga_{1-x}Mn_x As$, can be realistically represented within a generalized
Hubbard model involving multiple orbitals per site.\cite{magimp}

\subsection{Hartree-Fock ferromagnetic state}
In the Hartree-Fock (HF) approximation,
the interaction term reduces to a magnetic coupling
\begin{equation}
H_{\rm int}^{\rm HF} = - \sum_{I} 
\mbox{\boldmath $\sigma$}_I  . {\bf \Delta}_I  
\end{equation} 
with the local mean field ${\bf \Delta}_I$,
resembling the semi-classical limit of the exchange interaction
$ - \sum_I J \mbox{\boldmath $\sigma$}_I .{\bf S}_I $ in the spin-fermion model. 
Here the impurity spin operator
$\mbox{\boldmath $\sigma$}_I  =  \Psi_I ^\dagger [\mbox{\boldmath $\sigma$}] \Psi_I $
in terms of the fermionic field operator   
$\Psi_I =  \left ( \begin{array}{l} 
a_{I\uparrow} \\ a_{I\downarrow} \end{array} \right ) $.
The mean field ${\bf \Delta}_I$
is self-consistently determined from the ground-state expectation value: 
\begin{equation}
2{\bf \Delta}_I = \langle \mbox{\boldmath $\sigma$}_I  \rangle U \; .
\end{equation}

\begin{figure}[hbt]
\begin{center}
\vspace*{-70mm}
\hspace*{-38mm}
\psfig{figure=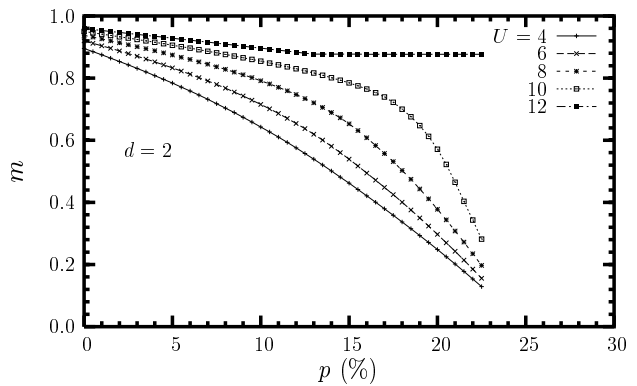,width=140mm}
\vspace*{-85mm}
\end{center}
\caption{Variation of impurity magnetization with hole concentration 
for the two-dimensional case $(x=25\%)$. With increasing $U$,
the impurity character of majority-spin states at the top of the band diminishes,
and the impurity moment is therefore less susceptible to hole doping.}
\end{figure}

\subsection{Sublattice-basis representation}
We consider ordered (superlattice) arrangements of magnetic impurities 
on square and cubic host lattices,
with impurity separations of $2a$ and $3a$. Translational
symmetry within the sublattice basis conveniently allows
Fourier transformation to momentum space.
For concreteness, we consider a square host lattice in the following,
with magnetic impurities placed at every other host site, 
corresponding to superlattice spacing $2a$ and impurity concentration $x=25\%$.
There are four sublattices, numbered $\alpha=1,2,3,4$,
corresponding to the four sites in the unit cell, as shown in Fig. 1.
We choose length and energy units such that
the lattice spacing $a=1$ and the hopping term $t=1$.

Without loss of generality, 
we assume a mean field ${\bf \Delta}_I = \Delta \hat{z}$ in the $z$ direction.
Fourier transformation within the four-sublattice basis
yields the HF Hamiltonian
\begin{equation}
H_{\rm HF}^{\sigma} = \sum_{\bf k} \Psi_{\bf k\sigma} ^\dagger \left [ 
\begin{array}{rccc} 
\epsilon_d - \sigma \Delta & \epsilon_{\bf k}^x & 0 & \epsilon_{\bf k}^y \\
\epsilon_{\bf k}^x & 0 & \epsilon_{\bf k}^y & 0 \\
0 & \epsilon_{\bf k}^y & 0 & \epsilon_{\bf k}^x \\
\epsilon_{\bf k}^y & 0 & \epsilon_{\bf k}^x & 0 \end{array} \right ]
\Psi_{\bf k\sigma}
\end{equation}
for spin $\sigma$, 
where $\epsilon_{\bf k}^x = 2t\cos k_x$
and $\epsilon_{\bf k}^y = 2t\cos k_y$ correspond to 
hopping terms in the $x$ and $y$ directions, respectively. 
Here the field operator 
$\Psi_{\bf k\sigma}= (a_{\bf k\sigma}^1 \; a_{\bf k\sigma}^2 \;
a_{\bf k\sigma}^3 \; a_{\bf k\sigma}^4)$
defines the sublattice basis,
where $a_{\bf k\sigma}^\alpha $ refers to the fermion operator
for sublattice index $\alpha$.
Generalization to other impurity concentrations and dimensions
is straightforward.
Hopping terms to the right (up) and left (down) do not connect
the same sublattice in general, 
yielding a Hermitian Hamiltonian matrix with hopping terms
$te^{\pm ik_x}$ etc.
The case of impurity spacing 2 is, however,
special as hopping terms in opposite directions
do connect the same sublattice, yielding a real symmetric
Hamiltonian matrix.

The HF Hamiltonian matrix is numerically diagonalized 
to obtain the four eigenvalues $E_{\bf k \sigma}^\mu$, 
corresponding to the four sub-bands $\mu=1,2,3,4$
in increasing order of energy.
The four-component eigenvectors $\phi_{\bf k \sigma}^{\mu\alpha}$ yield the 
amplitude on sublattice $\alpha$. 
Summing over occupied states yields the impurity magnetization
and the self-consistency condition
\begin{equation}
\frac{2\Delta}{U} = 
m(\Delta) = \langle \sigma_I ^z \rangle =
\sum^{E_{\bf k \sigma}^\mu < E_{\rm F}}_{\bf k, \mu}
(\phi_{\bf k \uparrow}^{\mu\alpha=1} )^2 -
(\phi_{\bf k \downarrow}^{\mu\alpha=1} )^2 \; .
\end{equation}
Variation of the impurity magnetization $m$ with doping concentration $p$
is shown in Fig. 2.

\begin{figure}
\begin{center}
\vspace*{-70mm}
\hspace*{-38mm}
\psfig{figure=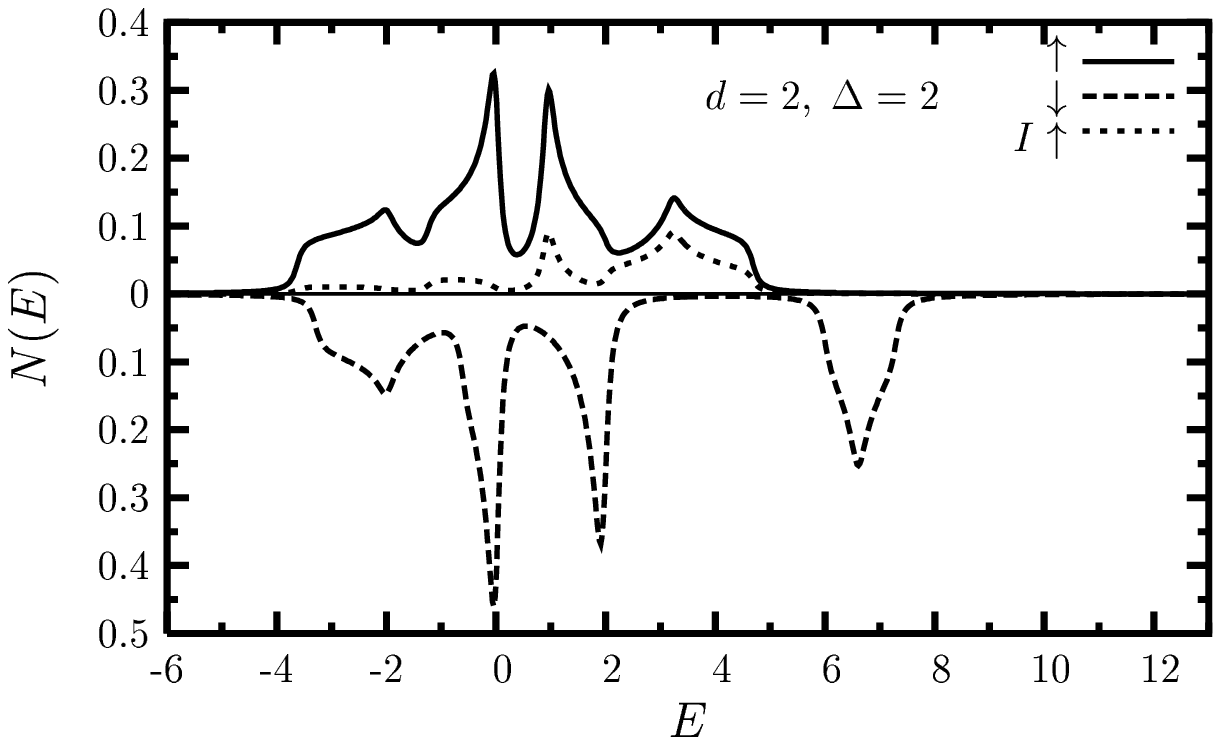,width=140mm}
\vspace{-75mm}
\vspace*{-77mm}
\hspace*{-38mm}
\psfig{figure=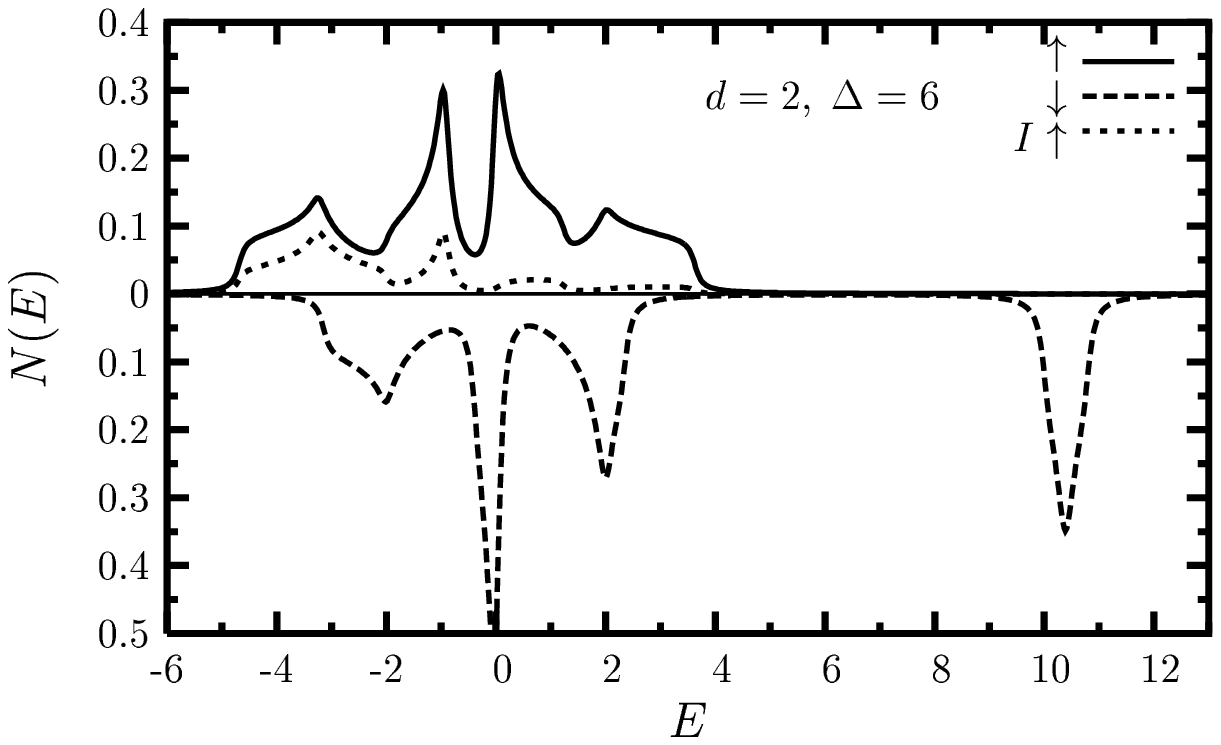,width=140mm}
\vspace{-85mm}
\end{center}
\caption{Total and impurity (I) density of states 
for the 2-d case $(x=25\%)$
for two values of the mean field $\Delta$,
showing the split-off impurity band and the impurity-induced sub-bands.}
\end{figure}

\begin{figure}
\begin{center}
\vspace*{-70mm}
\hspace*{-38mm}
\psfig{figure=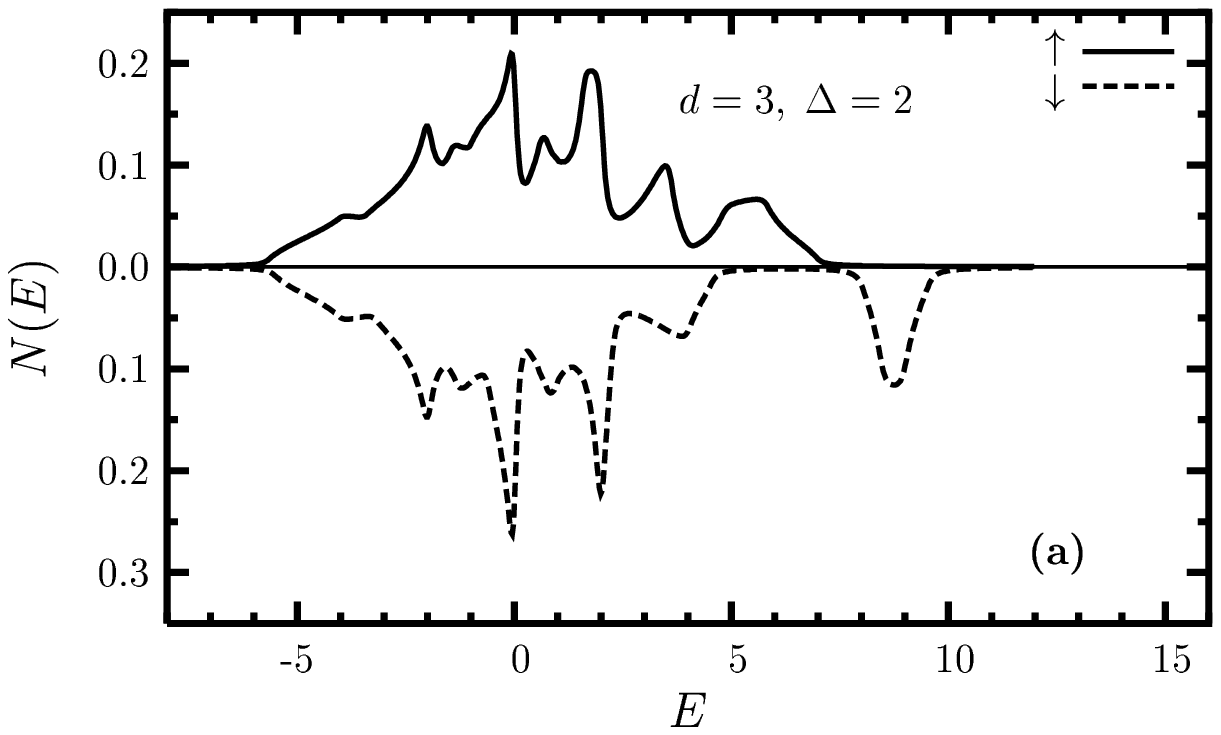,width=140mm}
\vspace{-75mm}
\vspace*{-77mm}
\hspace*{-38mm}
\psfig{figure=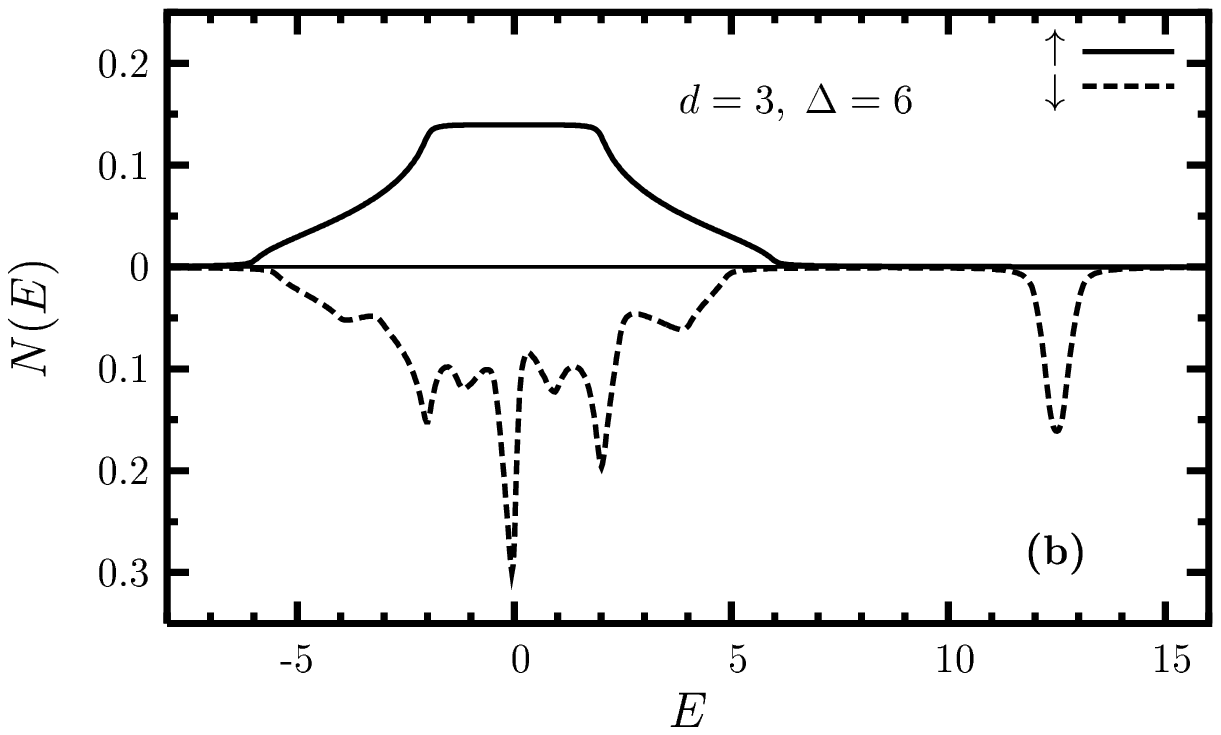,width=140mm}
\vspace{-85mm}
\end{center}
\caption{Density of states for the 3-d case $(x=12.5\%)$
for two values of the mean field $\Delta$.}
\end{figure}

\subsection{Quasiparticle spectrum}
Each magnetic impurity contributes two spin states.
The minority-spin impurity state (energy $\epsilon_d + \Delta$)
is pushed up by the local Coulomb repulsion,
and forms a split-off narrow impurity band due to the small overlap with
neighbouring impurity states.
The majority-spin impurity state (energy $\epsilon_d - \Delta$)
mixes with the host band states for $\epsilon_d - \Delta \ll W$,
and splits off on the low-energy side for large $\Delta$. 
Figures 3 and 4 show a comparison of the quasiparticle density of states
$N(E)$ for two $\Delta$ values for the 2-d and 3-d cases,
showing the four and eight sub-bands corresponding to the sublattices,
respectively.

For minority spin,
while the impurity band shifts up with increasing $\Delta$
and narrows due to decreasing effective impurity hopping
$t_{\rm eff} \sim t^2/(\epsilon_d + \Delta)$,
the host sub-bands remain unaffected
reflecting strong host-impurity decoupling.
The majority-spin sub-bands, however, are pulled down in energy by
the exchange interaction,
decreasing the energy separation between the two bands.
With increasing $\Delta$,
the impurity character of majority-spin states
at the top of the band diminishes,
as seen in Fig. 3.

In the undoped insulating state,
the Fermi energy lies in the gap between the top of the majority-spin
valence band and the minority-spin impurity band.
The empty impurity band ensures local-moment formation on each impurity site.
Carrier doping is introduced by adding holes
to the top of the majority-spin band.
The mean field $\Delta$ decreases with doping concentration $p$,
which shifts the majority-spin band to the right.
The Fermi energy therefore remains above the minority-spin valence
sub-band ($\mu=$3 or 7), yielding a finite Stoner gap for low doping.

\section{Transverse spin fluctuations}
Transverse spin fluctuations are gapless, low-energy excitations in 
the broken-symmetry state of magnetic systems possessing
continuous spin-rotation symmetry. 
Therefore, at low temperatures they play an important
role in diverse macroscopic properties
such as existence of long-range order, spin stiffness 
magnitude and temperature dependence of magnetization,
transition temperature, spin correlation etc.

We study the time-ordered, transverse spin propagator 
\begin{equation}
\chi_{ij}^{+-}(t-t') =  
i \langle \Psi_{\rm G} | T [ S_i ^+ (t) S_j ^- (t')]|\Psi_{\rm G}\rangle 
\end{equation}
in the the ferromagnetic ground state $|\Psi_G \rangle$,
involving the spin-raising ($S_i ^+$) and spin-lowering ($S_j ^-$) operators 
at sites $i$ and $j$.
In the random phase approximation (RPA), the magnon propagator
\begin{equation}
[\chi^{+-}({\bf q},\omega)]=
[\chi^{-+}({-\bf q},-\omega)]=\frac{[\chi^0({\bf q},\omega)]}
{{\bf 1}-[U][\chi^0({\bf q},\omega)]} 
\end{equation}
in ${\bf q},\omega$ space,
where the zeroth-order, particle-hole propagator
$[\chi^0({\bf q},\omega)]$ is obtained by integrating out the fermions
in the self-consistent broken-symmetry state. In the sublattice basis we have
\begin{eqnarray}
[\chi^0({\bf q},\omega)]_{\alpha\beta} &=& i\int \frac{d\omega'}{2\pi}
\sum_{\bf k}
[G^\uparrow({\bf k},\omega')]_{\alpha\beta}
[G^{\downarrow}({\bf k'},\omega'+\omega)]_{\beta\alpha} \nonumber \\
&=& 
\sum_{E_{\bf k \uparrow}^\mu < E_{\rm F}}
^{E_{{\bf k'} \downarrow}^\nu > E_{\rm F}}
\frac
{
\phi_{\bf k\uparrow}^{\mu\alpha}
\phi_{\bf k\uparrow}^{\mu\beta\ast}
\phi_{{\bf k'}\downarrow}^{\nu\beta}
\phi_{{\bf k'}\downarrow}^{\nu\alpha\ast}
}
{
E_{{\bf k'}\downarrow}^\nu - E_{\bf k \uparrow}^\mu - \omega
} \nonumber  \\
&+&
\sum_{E_{{\bf k'}\downarrow}^\nu < E_{\rm F}}
^{E_{\bf k\uparrow}^\mu > E_{\rm F}}
\frac
{
\phi_{{\bf k}\uparrow}^{\mu\alpha}
\phi_{{\bf k}\uparrow}^{\mu\beta\ast}
\phi_{{\bf k'}\downarrow}^{\nu\beta}
\phi_{{\bf k'}\downarrow}^{\nu\alpha\ast}
}
{
E_{{\bf k}\uparrow}^\mu - E_{{\bf k'}\downarrow}^\nu + \omega
} 
\; , 
\end{eqnarray}
where ${\bf k'} \equiv {\bf k+q}$.
In Eq. (7), the diagonal interaction matrix $[U]_{\alpha\alpha}=
U\delta_{\alpha 1}$
has non-zero element only on the impurity sublattice $\alpha=1$.
It is therefore convenient to write
\begin{equation}
[\chi^{+-}({\bf q},\omega)]=
\frac{1}{[A({\bf q},\omega)]} - \frac{1}{[U]} 
\end{equation}
in terms of a matrix $[A({\bf q},\omega)]=[U] - [U][\chi^0 ({\bf q},\omega)][U]$, 
which has non-vanishing matrix elements only on the impurity sublattice.
Magnon-mode energies $\omega_{\bf q}$ are therefore obtained
from the magnon pole condition
\begin{equation}
1-U\chi^0 ({\bf q},\omega_{\bf q}) = 0 \;  ,
\end{equation}
where $\chi^0 ({\bf q},\omega)$ represents the impurity-sublattice
matrix element of $[\chi^0 ({\bf q},\omega)]_{\alpha\beta}$.
In the ferromagnetic state, typically
$\chi^0 ({\bf q},\omega) = \frac{1}{U} - {\cal A}q^2 + {\cal B}\omega$
for small $q,\omega$, so that the magnon energy
$\omega_{\bf q}=({\cal A}/{\cal B})q^2$.
The coefficient ${\cal A}$ is a measure of the spin stiffness,
whose sign determines the stability of the
HF ferromagnetic state with respect to transverse fluctuations,
as discussed below.

\begin{figure}
\begin{center}
\vspace*{-75mm}
\hspace*{-38mm}
\psfig{figure=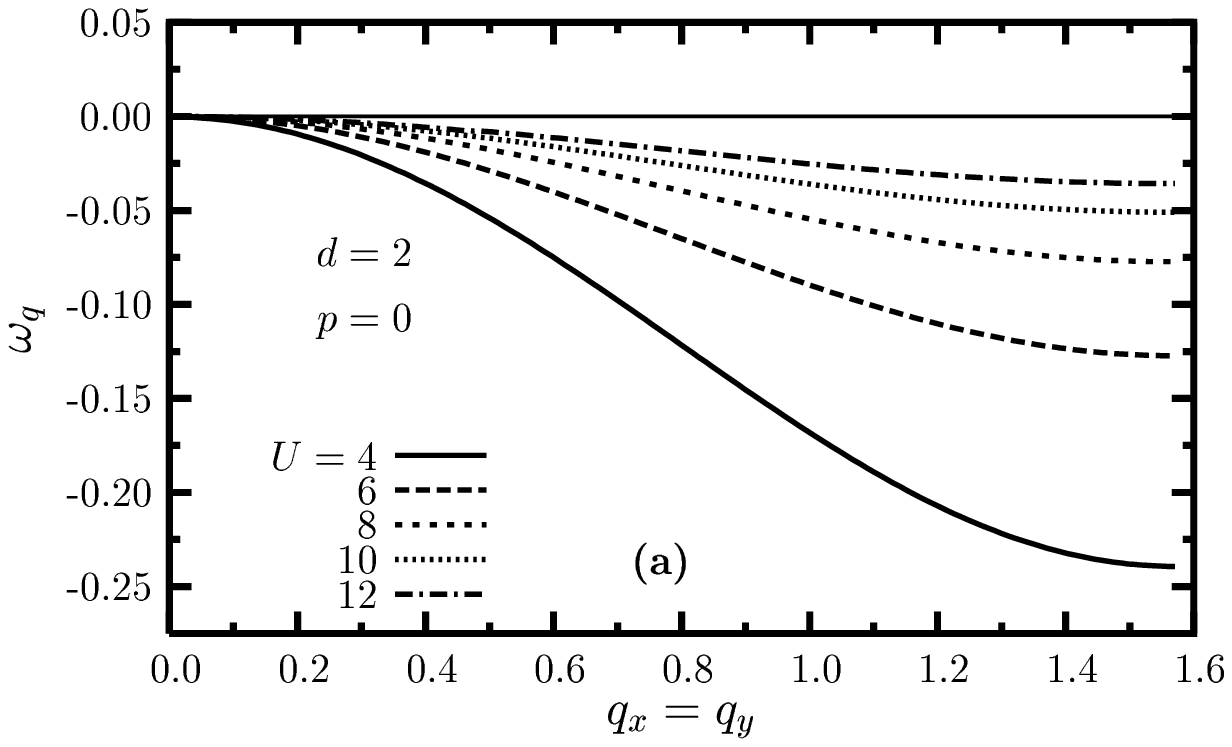,width=140mm}
\vspace*{-75mm}
\vspace*{-77mm}
\hspace*{-38mm}
\psfig{figure=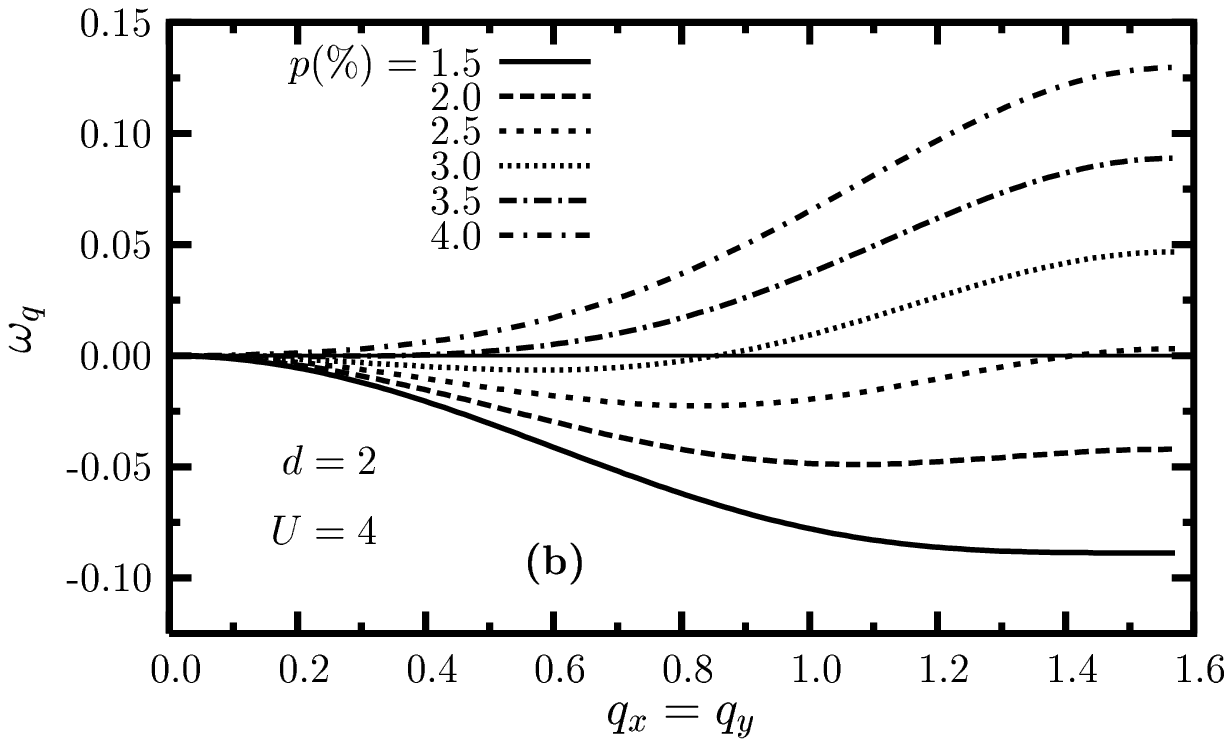,width=140mm}
\vspace*{-82mm}
\end{center}
\caption{(a) Negative magnon energy in the absence of hole doping 
indicates instability of the ferromagnetic state due to
antiferromagnetic spin couplings, which weaken with increasing $U$. 
(b) Stabilization of the ferromagnetic state with increasing hole concentration.}
\end {figure}

\subsection{Stability of the ferromagnetic state}
The undoped state, with filled majority-spin states
and an empty minority-spin impurity band, amounts to a half-filled case
with respect to magnetic impurities.
Virtual hopping between impurity sites therefore generates antiferromagnetic 
exchange interaction $t_{\rm eff}^2/U$ between neighbouring magnetic impurities,
resulting in antiferromagnetic ordering.
The carrier-mediated ferromagnetic couplings are absent, 
and the self-consistent ferromagnetic state is therefore unstable
and actually represents a saddle point.
The instability is reflected in a negative sign of the
coefficient ${\cal A}$ (spin stiffness),
which makes the transverse response eigenvalue $U\chi^0({\bf q}) >1$
and yields negative magnon energies [Fig. 5(a)].
With hole doping, ferromagnetic coupling strengthens,
long-wavelength magnon-mode energy changes sign,
and ferromagnetic ordering is stabilized 
at some critical hole concentration [Fig. 5(b)].
Thus the diluted Hubbard model is characterized by an AF - F transition 
at a finite doping concentration,
which is more prominent for small $U$.

A significant feature of Fig. 5(b) is the distinct behaviour
of long- and short-wavelength magnon modes.
Even in the unstable regime,
with negative-energy long-wavelength modes,
the zone-edge modes $(q_x,q_y \sim \pi/2)$ may have positive energy,
thereby giving a spurious indication of stability.
Incorporating only the zone-edge (Ising) excitations,
the dynamical mean-field theory is therefore insensitive
to the non-trivial long-wavelength behaviour arising from 
competing spin couplings.

\begin{figure}
\begin{center}
\vspace*{-75mm}
\hspace*{-38mm}
\psfig{figure=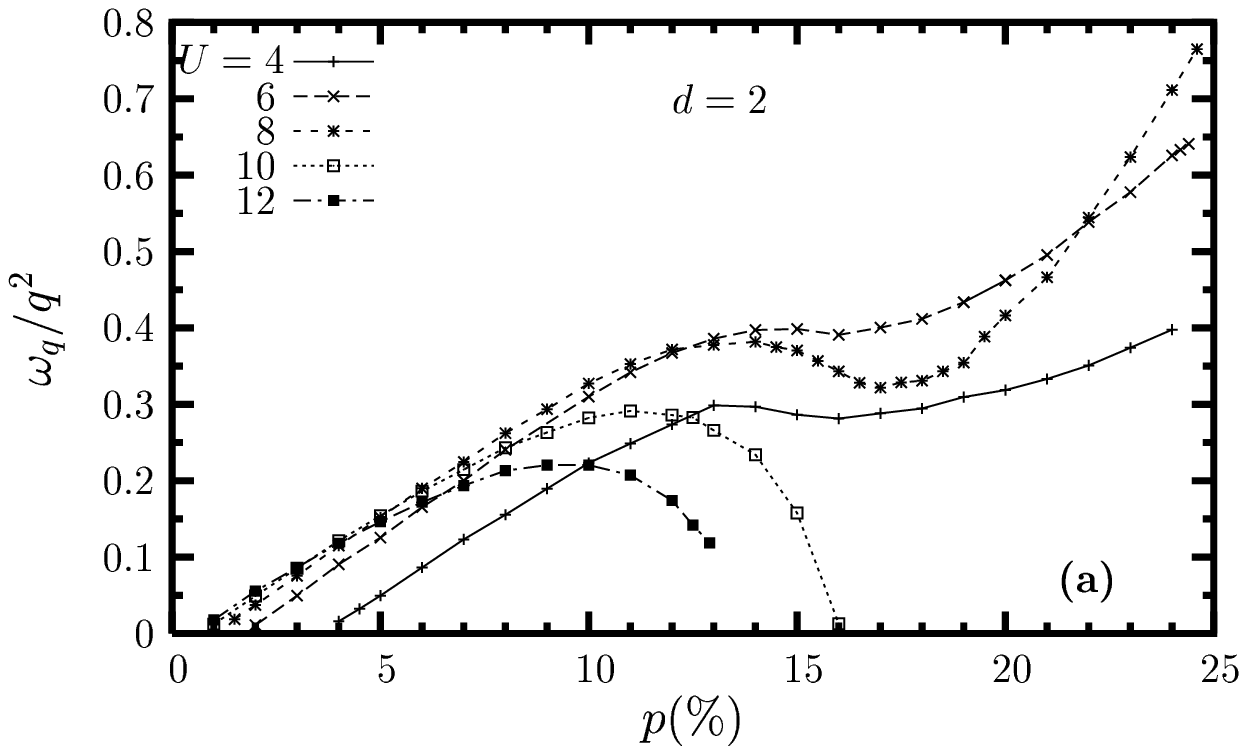,width=140mm}
\vspace*{-77mm}
\vspace*{-75mm}
\hspace*{-38mm}
\psfig{figure=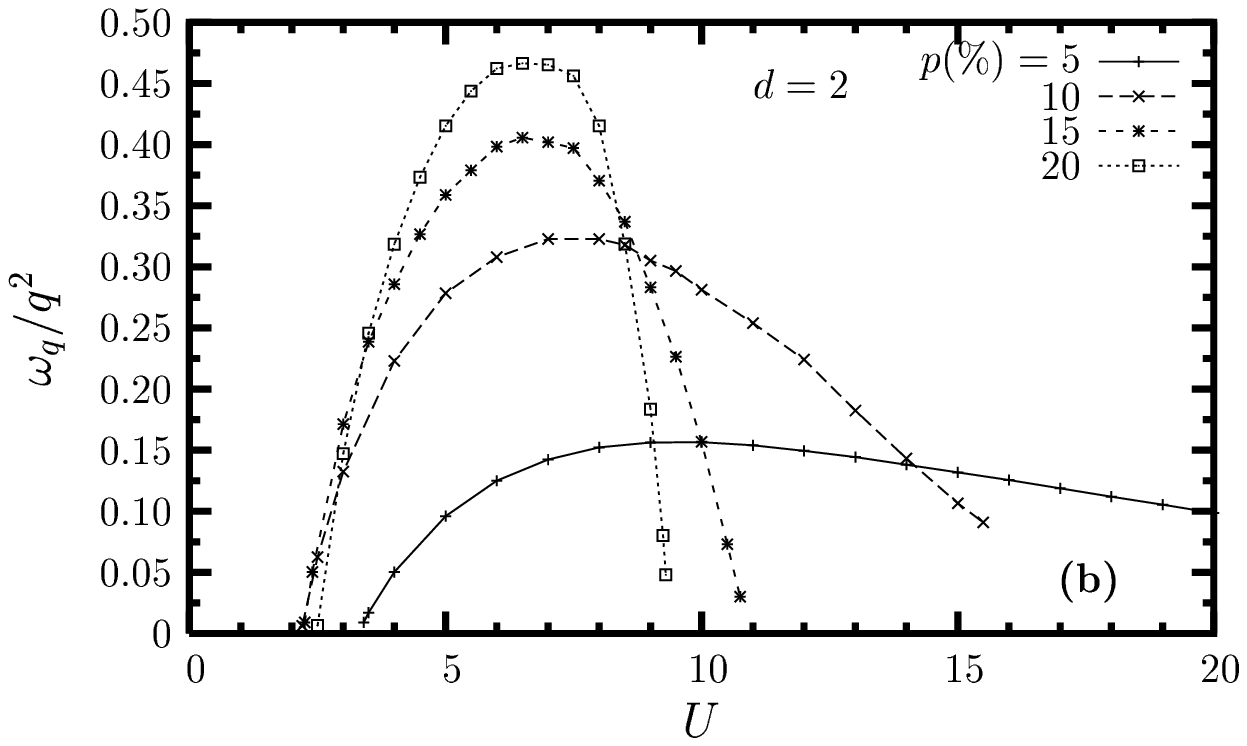,width=140mm}
\vspace{-82mm}
\end{center}
\caption{Variation of spin stiffness with (a) hole concentration 
and (b) interaction strength $U$, for the 2-d case $(x=25\%)$.}
\end{figure}

\subsection{Spin stiffness}
The long-wavelength magnon-mode energy $\omega_{\bf q}$
and spin stiffness $\omega_{\bf q}/q^2$
provide a composite measure of impurity spin couplings in the
carrier-mediated ferromagnetic state.
Figures 6 and 7 show the behaviour of spin stiffness
with doping concentration $p$ and interaction strength $U$
for the 2-d and 3-d cases with impurity spacing $2$.
A minimum hole concentration to stabilize the ferromagnetic state
is clearly seen, especially for low $U$, when the competing antiferromagnetic
spin couplings are relatively stronger.
The increase in spin stiffness with hole concentration
exhibits the essence of carrier-mediated ferromagnetism.
The initial increase is nearly linear in two dimensions
and distinctly sublinear in three dimensions.
The spin stiffness exhibits an optimization with respect to both $p$ and $U$,
which can be understood qualitatively within the generalized RKKY theory
in terms of a competition between
increasing magnitude of carrier-spin polarization
and increasing rapidity of its oscillation.

Similar behaviour of spin stiffness is seen (Fig. 8)
when impurity spacing is increased to 3 in the 2-d case.
The peak spin stiffness is clearly reduced,
reflecting the weakening of spin couplings with increased
impurity separation. 
However, in the low doping regime,
the spin stiffness for $x=1/9$ is actually greater than that for $x=1/4$
(see inset).
As the stiffness must eventually decrease with increasing impurity separation,
this interestingly shows an optimization behaviour with respect to
impurity concentration $x$ as well.

The anomalous increase at higher $p$,
seen in Figs. 6-8 for lower $U$ values,
is due to a competition between two particle-hole processes, as discussed below.
Figure 9 shows the doping dependence of
$U^2[\chi^0(0)-\chi^0({\bf q})]/q^2$ for the two particle-hole processes
in Eq. (8), labeled as 1 and 2, which provide measures of the
corresponding spin-stiffness contributions. 
Positive (negative) value indicates dominant ferromagnetic (antiferromagnetic)
spin coupling. Process 2, which is activated only upon doping in the
majority-spin band, is seen to be dominantly ferromagnetic in the
low-doping regime and exhibits the typical RKKY peak behaviour
before turning negative due to the oscillating carrier-spin polarization.
However, process 1, which is weakly antiferromagnetic for $p\approx 0$,
changes sign with doping, and becomes increasingly ferromagnetic;
it is the sharp increase seen at higher doping which causes the distinct
minimum feature in the total spin stiffness.
This feature is more pronounced at lower $U$ because the sharply reduced
impurity moment and $\Delta$ at higher doping (see Fig. 2),
cause the impurity states to be more extended,
thus enhancing the process 1 stiffness. 

\begin{figure}
\begin{center}
\vspace*{-75mm}
\hspace*{-38mm}
\psfig{figure=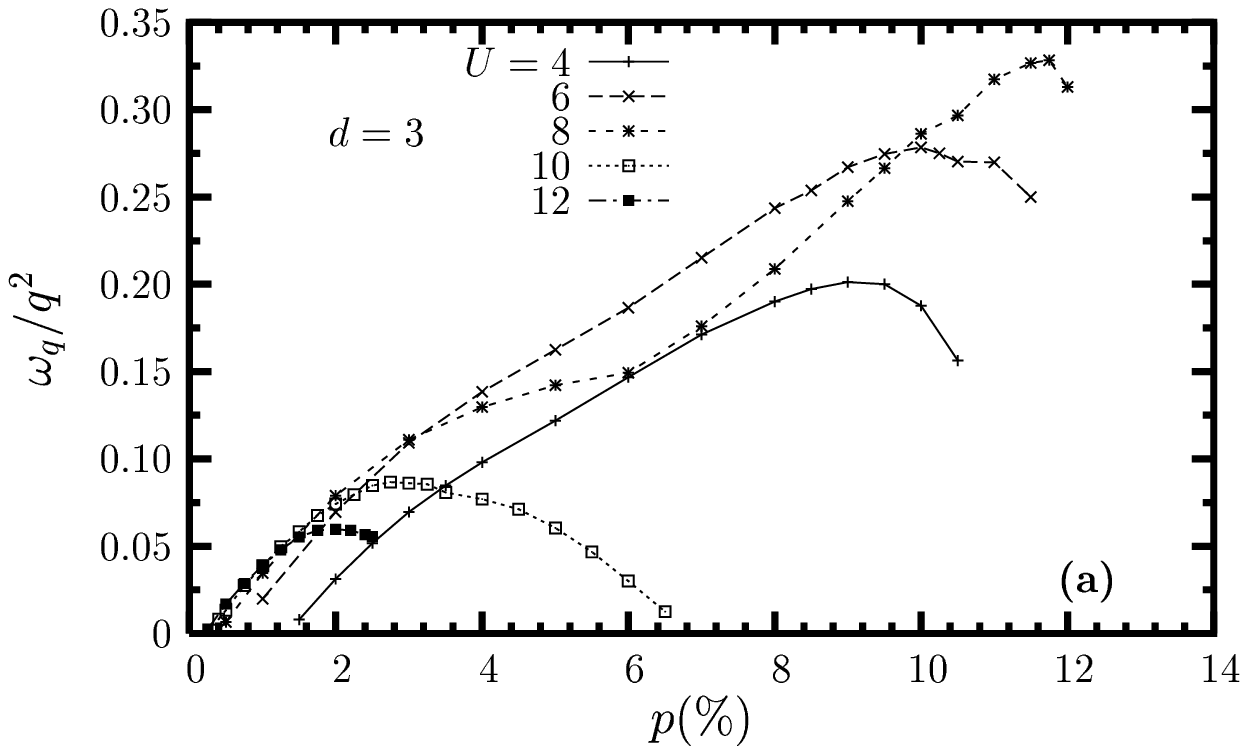,width=140mm}
\vspace*{-75mm}
\vspace*{-77mm}
\hspace*{-38mm}
\psfig{figure=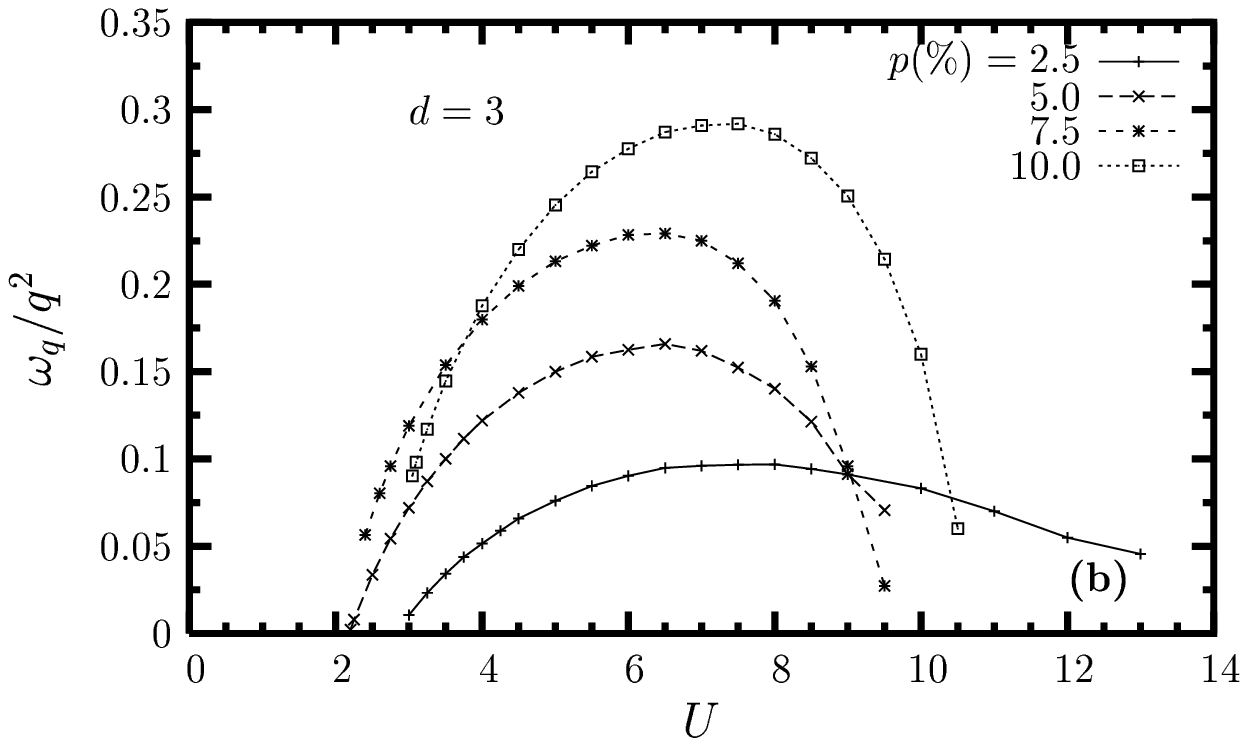,width=140mm}
\vspace{-82mm}
\end{center}
\caption{Variation of spin stiffness with  (a) hole concentration 
and (b) interaction strength $U$, for the 3-d case $(x=12.5\%)$.}
\end{figure}

\subsection{Transition temperature}
Determination of the
magnon spectrum in the carrier-mediated ferromagnetic state
allows an estimation of the transition temperature $T_c$ 
for the diluted Hubbard model in three dimensions.
As the ferromagnetic state is characterized by small spin stiffness
due to strong competition between
ferromagnetic and antiferromagnetic spin couplings,
the dominant contribution to reduction in magnetization
is from the thermal excitation of long-wavelength magnon modes.
Therefore, the transition temperature $T_c$ can be estimated
in terms of an equivalent Heisenberg model with matching
spin stiffness.

\begin{figure}
\begin{center}
\vspace*{-67mm}
\hspace*{-38mm}
\psfig{figure=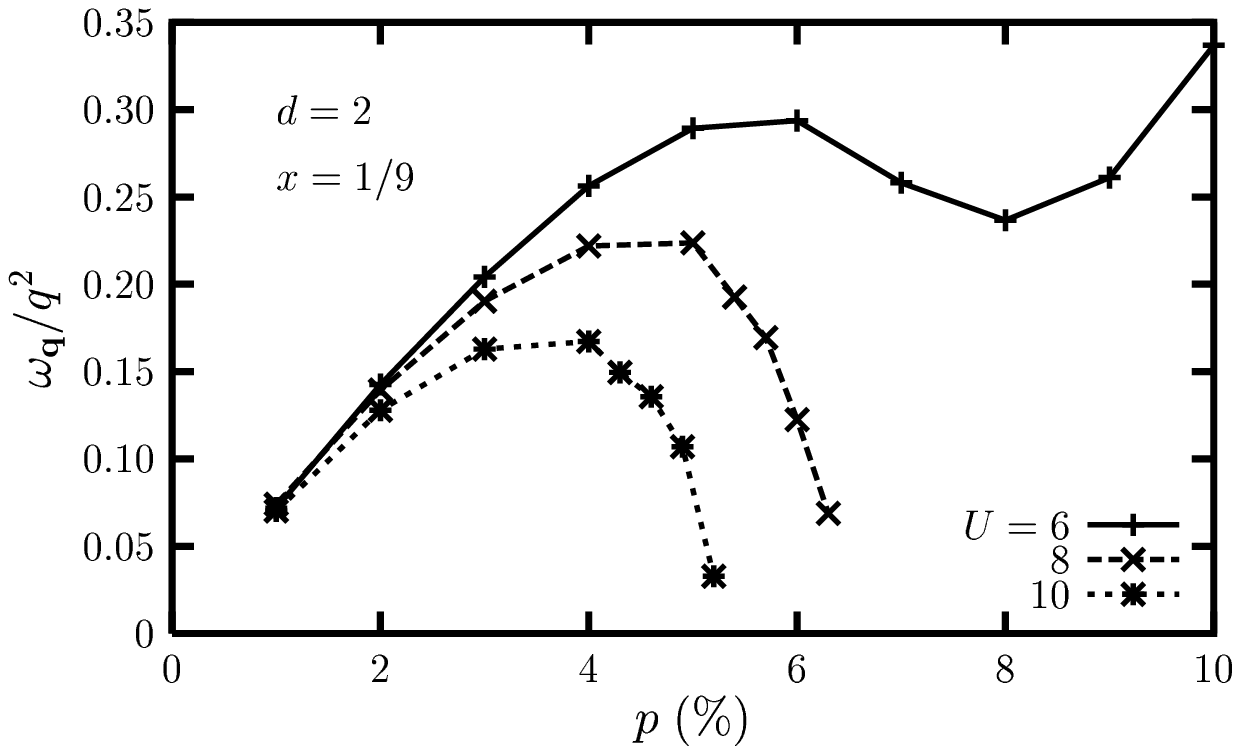,width=140mm}
\vspace{-82mm}
\end{center}
\caption{Increasing the impurity spacing to 3 results in
qualitatively similar but reduced peak spin stiffness.
Inset shows comparison of the $x=1/4$ and $x=1/9$ cases for $U=8$.}
\vspace*{-76mm}
\hspace*{40mm}
\psfig{figure=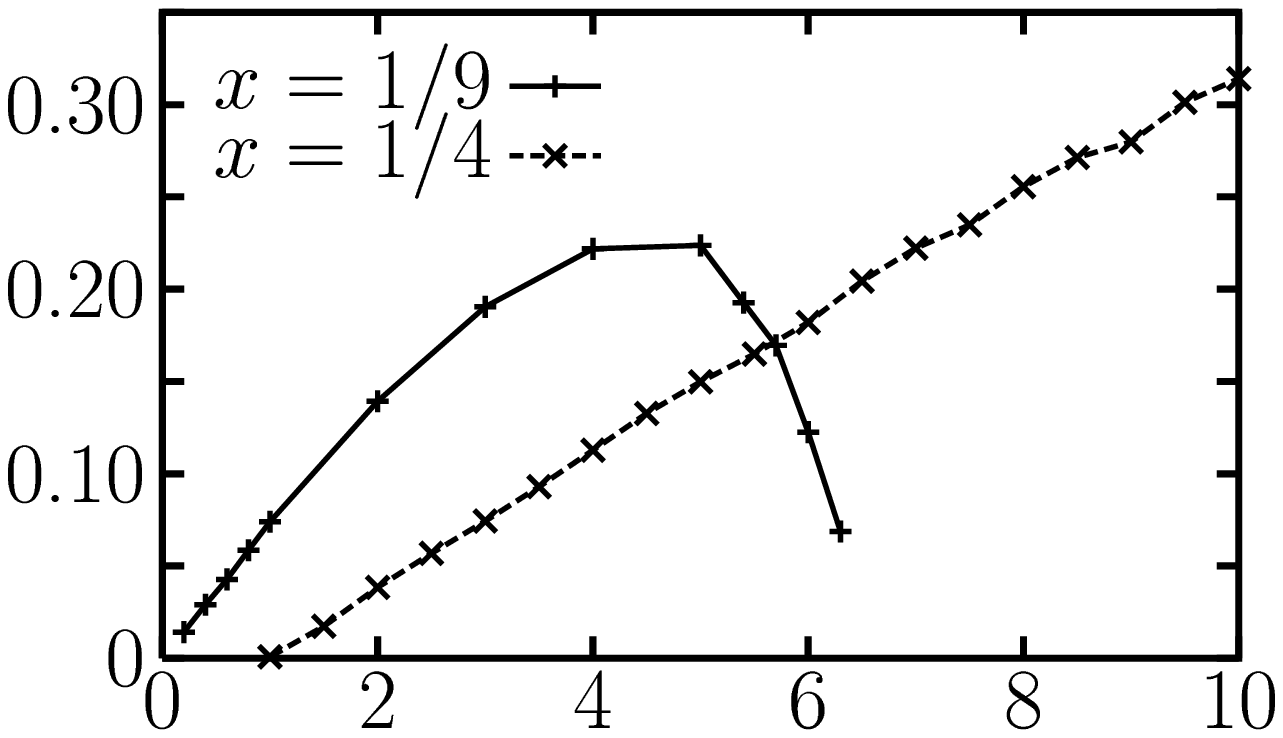,width=45mm}
\end{figure}

\begin{figure}
\begin{center}
\vspace*{-55mm}
\hspace*{-38mm}
\psfig{figure=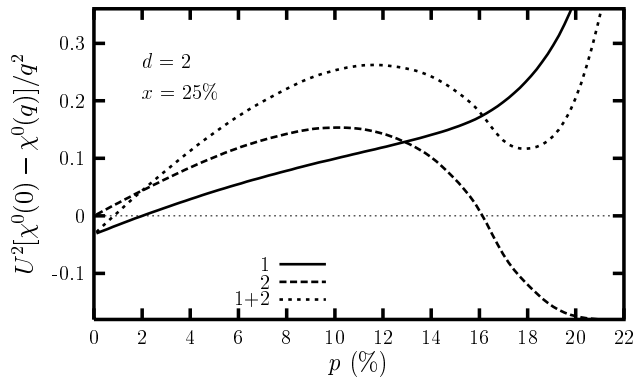,width=140mm}
\vspace{-82mm}
\end{center}
\caption{Spin-stiffness contributions corresponding to the two
particle-hole processes in Eq. (8),
showing the distinct minimum feature and the anomalous increase
arising from their competition.
Here $U=9$.}
\end{figure}

For a nearest-neighbour, spin-$S$ quantum Heisenberg model with
interaction energy $J$ on a cubic lattice (coordination number $z=6$),
the magnon energy 
\begin{equation}
\omega_{\bf q}=zJS(1-\gamma_{\bf q}),
\end{equation}
where $\gamma_{\bf q} = (\cos 2q_x + \cos 2q_y + \cos 2q_z)/3$,
corresponding to the magnetic lattice spacing $2$.
Indeed, the magnon energy is maximum for $q_x = q_y = q_z = \pi/2$,
as also seen in Fig. 5 for the 2-d case. 
Considering the small $q$ limit of Eq. (11),
and matching the spin stiffness,
we obtain $D=\omega_{\bf q}/q^2 = 4JS =2J$,
where we have set $S=1/2$ for the diluted Hubbard model
having a single magnetic orbital per site.

\begin{figure}
\begin{center}
\vspace*{-67mm}
\hspace*{-38mm}
\psfig{figure=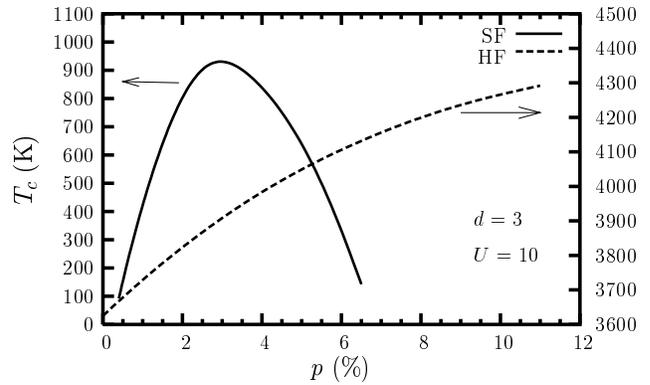,width=140mm}
\vspace*{-83mm}
\end{center}
\caption{Comparison of the transition temperature $T_c$ 
obtained from the renormalized spin-fluctuation theory (SF) 
with the HF result (temperature scale on the right),
highlighting the order-of-magnitude
difference between the spin-ordering and moment-melting temperatures.
Here $x=12.5\%$.}
\end{figure}

For the spin-$S$ Heisenberg model, the transition temperature
\begin{equation}
T_c ^{\rm SF}= \frac{zJ S(S+1) }{3} f_{\rm SF} ^{-1}  = D S(S+1) f_{\rm SF} ^{-1} 
\end{equation}
within the renormalized spin-fluctuation theory 
is somewhat lower than the mean-field value $T_c^{\rm MF} = zJ S (S+1) /3$
due to the spin-fluctuation factor
$f_{\rm SF} = \sum_{\bf q} (1-\gamma_{\bf q})^{-1}$,
which is approximately 1.5 for the cubic lattice.
Extrapolating to spin-$S$ magnetic impurities in the diluted Hubbard model,
represented by multiple magnetic orbitals per site,\cite{magimp}
with the same equivalent interaction energy $J = D/2$ as obtained above, 
the transition temperature reduces to the expression in Eq. (12).
Taking $S=5/2$ for Mn impurities,
and a realistic heavy-hole bandwidth $W=12t=2$ eV,
the transition temperature evaluated from Eq. (12) is shown in Fig. 8.
Also shown for comparison is the HF result obtained from a
finite-temperature analysis of the self-consistency condition.
The HF $T_c$ is naturally an order of magnitude higher
as it really corresponds to a moment-melting temperature.

At the HF level, the impurity moment vanishes at a temperature
$T \sim \Delta$, the minority-spin impurity-band energy 
relative to the Fermi energy.
The marginal increase seen in $T_c^{\rm HF}$ with doping is simply due to
the decreasing Fermi energy, whereas the impurity-band
energy remains essentially unchanged.

Similar estimations of $T_c$ from magnon energy were recently carried out
for the ferromagnetic Kondo lattice model\cite{vogt}
and in the context of DMS,\cite{koenig,pareek,prelim,dms_nolting} 
as a function of carrier concentration $p$.
Optimization behaviour was found in these cases as well,
with $T_c$ increasing with $p$ upto a maximum,
followed by a monotonic decrease.

\section{Conclusions}
Carrier-induced ferromagnetism was investigated in the diluted Hubbard model
for ordered impurity arrangements.
Momentum-space representation within the sublattice basis permitted a
detailed study with respect to carrier doping, impurity spacing,
electron correlation, and wave vector. 
The delicate competition between spin couplings results in an
AF-F quantum phase transition in the low-doping regime.
Competition between the increasing magnitude
and increasing rapidity of oscillation of the carrier-spin polarization
was observed, yielding a characteristic optimization of the spin stiffness with 
doping concentration $p$ and interaction strength $U$.
Surprisingly, in the low doping regime,
the spin stiffness was found to actually increase 
when the impurity separation increased from 2 to 3 in the 2-d case,
indicating an interesting optimization with respect to 
impurity concentration as well.
In addition, the itinerant ferromagnetic state exhibits
a subtle interplay of impurity moment reduction, impurity character of doped states, 
and competing particle-hole processes in the carrier-spin polarization,
resulting in a distinct minimum and an anomalous increase in the spin stiffness
at higher doping concentration.

For the cubic impurity arrangement, doping dependence of the ferromagnetic
transition temperature $T_c$ was calculated within the renormalized
spin-fluctuation theory from the magnon spectrum,
and compared with the HF result so as to highlight the difference between
spin-ordering and moment-melting temperatures.
For $S=5/2$ and a realistic heavy-hole bandwidth of 2 eV,
the peak calculated $T_c$ of 960 K at $p \approx 3 \%$ is in good agreement with the
highest reported $T_c$ value of 940 K observed in $\rm Ga_{1-x} Mn_x N$.\cite{GaMnN3}

While our main objective was to examine the novel ferromagnetism
in the diluted Hubbard model,
several features of this itinerant ferromagnetic state are of relevance
to DMS systems as well. For example, impurity bands formed by overlap
of Mn d orbitals are prominent features in density functional calculations
within the local spin density approximation (LSDA), and a characteristic
dependence of $T_c$ on Mn concentration is obtained
for impurity bands formed within the gap,
as is the case for $\rm Ga_{1-x} Mn_x N$.\cite{sato}
Furthermore, correlation effects have also been examined recently
within the LSDA+U approach\cite{sanyal} to obtain better agreement of the Mn d-DOS peak
position with recent photoemission experiments.

\section{Acknowledgement}
Helpful discussions with Indra Dasgupta are gratefully acknowledged.

\end{document}